# Policy for Access

## Framing the Question[*]

David Allen[**]

ABSTRACT

Five years after the '96 Telecommunications Act, we still find precious little local facilities-based competition. The infrastructure for local broadband access is also substantially behind expectations. In response there are calls in Congress and even from the FCC for new legislation to "free the Bells." However, the same ideology drove policy, not just five years ago, but also almost twenty years back with the first modern push for "freedom," namely divestiture.

How might we frame the question of policy for local access to engender a more fruitful approach? The starting point for *this* analysis is the network—not bits and bytes, but the human network. With the human network as starting point, the unit of analysis is the community—specifically, the individual in a tension with community. There are two core ideas.

The first takes a behavioral approach to the economics. The insight here is that each of us seems to come innately equipped with effortless temporal alternation between two key opposites, competition and consensus. Then competition as a key component of policy can be joined by what amounts to its opposite, but in a temporal succession. For this novel policy frame the marketplace operates in a tension with community hierarchy—the relative share between beneficial chaos and order, in economic affairs, becomes explicit.

If the first main idea provides a conceptual base for open source, the second core idea distinguishes open source from open design—the focus turns, that is, to the information 'frontier' we push forward. While the first policy innovation mandates organizational behavior, the second notes that the end product of such a policy cycle is actually a *choice* between open and integrated design. The trade is between 'integrated' design for higher performance now and 'open' design for more innovation down the road and so better performance *later*.

The resulting policy frame for access is worked out in the detailed, concrete steps of an extended thought experiment. A small town setting (Concord, Massachusetts) grounds the discussion in the real world. The purpose overall is to stimulate new thinking

---

which may break out of the conundrum where periodic rounds to legislate 'freedom' produce the opposite, recursively. The ultimate aim is better fit between our analytically-driven expectations and economic outcomes.

# Policy for Access
## Framing the Question[*]

### David Allen[**]

This chapter aims to frame the question of policy for access. Though the approach is unaccustomed, the question of access is one of the more subtle and challenging. The hope is to provoke new insight. In the course of erecting a scaffold for the frame, the chapter also investigates some of the surprising implications that arise.

The starting point for the analysis is the network—not bits, bytes, and the whole panoply of physical nets, but the human network. This is the collection of personal ties that you and others feel. While the sense of a personal tie is only in each person's head, the feeling is shared (if each person may see it slightly differently from everyone else). The result is a 'network of shared ties,' one of the more compelling forces in human affairs.

Are you skeptical that such human networks play their part or even exist? One has only to look to the 2000 U.S. presidential election for an apt case. Two camps of dogged partisans—two networks, each with shared ties—each banded together in common cause against the other, with the presidency the prize that only one would take away.

## Roadmap

Discussion begins with several tenets for a new access policy. The policy is brought to some life and given a workout with a thought experiment. Two bedrocks of the analysis surface in the process. The chapter concludes asking who is us and will the new access policy work.

## Analysis for a policy

THE UNIT OF ANALYSIS—INDIVIDUAL IN COMMUNITY

With the human network as starting point, the unit of analysis is the community. Or more accurately, the unit is the individual set in the community—the individual in a tension with community. Then policy for access to the wires and ether is easy. The physical network clearly serves this community. So access policy is simple: for the network to do its job, everyone must get onto the network—if they are in the community,

NB: Open links to access URL's, if viewing electronically.

[*] Paper available for download at http://www.davidallen.org/papers/Policy for Access.pdf and http://www.davidallen.org/papers/Policy for Access A4.pdf (there are spaces in the URL's).
[**] Co-Principal, World Collaboration CPR; Co-Editor, Information Economics and Policy; 316 Heaths Bridge Road, Concord, MA 01742, USA; +1 978 287 0433 / +1 978 287 0434 – fax; David_Allen_AB63@post.harvard.edu; www.davidallen.org , www.worldcollaboration.org [preview site available shortly].

anyway. In this brave new network, there is no occasion for discussion about how to provide access for the 'edge,' for those who in another world would be beyond the margin of the physical connections. With this policy everyone gets a connection, period. That is a foundation tenet of such an analysis.

This begs of course a very large question: what defines community. The chapter will turn to confront this, but only penultimately when new tools are in hand.

Yes, when there is new technology for the connection, such as broadband, there will be issues of transition, from old to new. Who will enjoy the fruits first, and who will have to wait for rollout? But in this policy with community as the unit of analysis, the rollout may start with the farm family rather than city dwellers. After all, the impact on quality of life may be greatest for community members who are relatively more isolated by geography. But whether some may not enjoy the same connection, after a measured rollout, is *not* a question here—everyone connects in this policy

COROLLARY POLICY TENETS

For everyone to connect, a corollary tenet of policy naturally holds that the physical network must be affordable. In the United States, physical networks represented by the highway and postal systems have long used such an approach. In fact we speak of postalized, that is distance-insensitive, pricing. Domestic long distance prices in the United States have been postalized for several years now; a few international rates, such as to the United Kingdom, are even approaching the same level. Also, policy makers in the United States' trading partners have envied U.S. local telephone flat rates. Flat rate is another facet of affordability.

Affordability may require subsidies. Half a century ago after the Second War, income distribution in the United States, for an example, was remarkable for being evenly spread to a quite broad middle class. In a stunning and dramatic shift over those fifty years, equally remarkable today is one of the most skewed income distributions among major industrialized countries. Now affordability for the network connection likely means some modest wealth transfer. Only then, it seems likely, will the now-marginalized couple of lower deciles make the connection.

As a third tenet, policy for this access-to-everyone must be dynamic. As technology continues to ratchet up the bar for broadband, the policy must provide for serving everyone with the same quality of connection. How does the network connect the whole community at the same speed, all the while the definition of 'fast' continues to change under their feet? Rather than fix upon some bandwidth number that will define broadband for all time, a dynamic access policy repeats a sequence: 'freeze the definition,' then 'change it'; 'freeze the definition,' then 'change it'; 'freeze the definition,' then 'change it' …

Clearly this calls for a significant change in mindset. Policy looks for certainty by trying to lock onto some fixed number. But in a dynamic world where change is endemic, the policy vision must be able to accommodate that ongoing change. Instead of a fixed number, access policy is a time series of escalating numbers, each fixed for a period unknown ex ante and where the height or duration of future 'fixings' is also unknown. The policy community will do well not to underestimate the challenge of winning minds and mindsets to such a new dynamic schema.

With change an explicit part of policy, change is conscious rather than left to some unseen hand. Just how much—and especially when—to bring order and how much and when to unleash chaos, for access policy, is a subject that unfolds with the analysis.

Perhaps the most surprising implication for access policy is the mechanism. The present mantra, the Holy Grail, exported around the world via U.S. trade policy, for telecommunications and otherwise, is 'competition.' Competition is crucial, and it is also difficult indeed to sustain. But with community as the starting point—the network of ties among people—another mechanism is equally important. 'Concerted action' must operate directly alongside competition. With neither one nor the other greater nor lesser than the other, concerted action and competition are twinned together. How this seeming contradiction may resolve also unfolds below.

Such a radical departure from today's received wisdom requires, at a minimum, a conceptual framework if it is one day to have prospect for consideration. As just noted, a sea change in mindset would be in the offing. Such a shift will also look, among others, for suitable conceptual anchors. Those who want to step back and work from that next level—a conceptual system for this analysis—will find a start at (Allen 2000, available as a 228K download at <http://www.davidallen.org/papers/Liberal Evolution.pdf>. Please note that the references useful to this chapter's line of analysis will also be found there.) The conceptual treatment there is particularly concerned with preserving the precious and hardwon place for individual choice through competition, at the same time recognizing and evolving to accommodate the inseparably social character of economic and human endeavor.

**A thought experiment—policy tenets in action**

To look more closely at the particulars of this policy for access as it would actually be used, imagine a thought experiment. Look at a community as it formulates access policy.

This thought experiment focuses on Concord, Massachusetts, in the United States. The author is more than familiar with Concord; but its setting and particularly its recent history on the question of access also lend themselves well, to demonstrate policy in action.

To give a sense for the place: Concord is a community of 17,000 about twenty-five miles west of Boston, set on the Massachusetts plain at the confluence of small local rivers. With a core cluster of homes in an inviting New England town center, many of the residents also live in pastoral settings a few miles out from the center. Though a typical suburban bedroom community, with some high tech industry, the town has a dimension beyond usual suburbia; for whatever reason this tends also to reflect in community life of the town: Despite its small size, Concord attracts a million visitors a year for its history, both as the place where shots were first fired, igniting the American Revolution against King George, in the eighteenth century and also as incubator for the Transcendentalist writers of the mid–nineteenth century, such as Emerson and Thoreau.[1]

THE EXPERIMENT OPENS

As this thought experiment opens, the town will be at some point in the cycle of 'freeze, change'; freeze, change'; freeze, change' … for purposes of access policy. To commence the thought experiment, start at that moment when broadband options such as cable modems and DSL begin to appear. That was the later part of the 1990's; then the 'freeze' that precedes had fixed upon voice telephony and 56k phone modems.

Now facing new technological options the town, in this thought experiment, is at the threshold of a next cycle.[2] First it must 'change' its access arrangements, then once again 'freeze' them. 'Freeze' is necessary because networks only work if there is coherence across them. (Recall that in this access policy nobody is left out and everyone enjoys the same quality of network—so here network coherence extends even to equal bandwidth, and so forth.) 'Change' breaks that coherence, but in the interests of establishing what next to freeze.

This sequence allows of course both innovation to enter the network *and* the network to serve its community faithfully. This cycle between innovation and standardization is the basic building block for a dynamic understanding of technical change in a network.

To succeed at the first half of the cycle—'change'—the town needs to break down into parts and conduct its own set of separate experiments (and definitely more than mind experiments!) For the second half—'freeze'—it will come back together and confer, pick a best choice and freeze that in place. Until the next cycle …

To check this thought experiment, midstream, against the real world: Note that this path describes concisely the steps suggested by the Harvard Information Infrastructure Project for the state of California. The state confronted similar questions. The recommendation was first to conduct separate experiments with different access technologies (each set in its own locale across the state), then to assess the results and make a choice. The town of Concord did itself appoint a citizen committee of

---

[1] Besides serving as the Chair, on and off over the decade of the 1990's, for Concord's citizen and volunteer Cable TV Committee, the author organized a later citizen committee to investigate fiber-to-the-home as a possible municipal enterprise for Concord.

[2] For the thought experiment—and for that matter most the other writing on the subject—the cycle begins with 'change' (or 'innovation'), then moves to 'freeze' (or 'standardization'). The whole point is the dynamics, so that in a real sense there is no starting place as such and either ordering of the sequence is equally descriptive.

knowledgeable residents charged with investigating the options, particularly the possibility for fiber-to-the-home provided as a town utility. A neighboring town, Shrewsbury, Massachusetts, which is no more than about twice the population of Concord, actually did conduct its own experiment. Shrewsbury installed a fiber-to-the-home trial system, to test the possibility for a town utility. Recently, in 2000, there has been a burgeoning of similar experiments/announcements in North America, including places small and large, such as Chicago. [More about the rest of the world below. The CANARIE-NEWS list from Bill St. Arnaud in Canada is an excellent source for staying abreast of what are rapid developments.³]

TIME FOR A 'CHANGE'

To launch the thought experiment, look at the step immediately ahead for the town: the 'change' half of the cycle. New ideas need to be tried out, to test possibilities. This is the time for classical competition among ideas, as put forward by individual innovators and entrepreneurs. In this first half of the cycle, rather than behave as a coherent group, the community devolves to its elements in pursuit of good new ideas and competitive tests of those proposals. This is the occasion to let loose the spirit of creative chaos, in other words this is the time for freedom of action and for the individual to innovate—the marketplace.

In the thought experiment, imagine that the town fathers, and mothers, with assembled expert committees, investigate and try out the likes of cable modems, DSL, wireless local loop, direct broadcast satellite, and fiber-to-the-home. (For now, leave in suspension the question of a small town's capacity to test several different possible systems, when the relatively large investment required is typically one barrier to experimenting with communications networks—even if a small town like Shrewsbury came close to doing just that. Size of community, as pivotal variable, will reappear with the question of who is in the community.)⁴

*Now*, TIME TO 'FREEZE'

When the experimentation phase comes to a close for Concord (*in* the thought experiment), the town shifts gears and moves to the second half of the cycle: it will choose, then 'freeze' a new access method for the town, until the next cycle. Now the

---

³ To subscribe to CANARIE send email to: majordomo@canarie.ca. In the body of the email put:
   subscribe testnet
   end
⁴ The Concord expert committee on town communications infrastructure ultimately, if informally, decided in favor of recommending that Concord pursue fiber-to-the-home through its excellent town-owned electric utility. The model included town-provided conduit, with service providers competing across the glass. As it turned out, there was a parallel need to spend upwards of $50 million on school renovation; that is a large sum in a small town. Fiscal overhang from the school building project made unrealistic, for the moment anyway, committing the electorate to perhaps another $10 million on a new fiber system. Shrewsbury trialled fiber-to-the-home early enough in the technology development cycle that the results were not satisfactory. A trial today, with the optoelectronic interface now proved in costwise, might well come out favorable; at least that is the evidence from burgeoning fiber announcements across North America.

Concord citizens doff their innovator/entrepreneur hats and retake their accustomed places in the informal town hierarchy that is characteristic of New England town democracies. They must—each from their accustomed perches in the informal hierarchy, but deliberating together by means of that hierarchy's well-worn if informal processes—mull the results, perhaps meld together some of the proposed ideas with other ideas in the trials, and pick a winner. *Now* is the occasion to reap the fruits of concerted action—now is the moment to utilize the machinery of hierarchy and consolidate the benefits that order in economic behavior can bring.

*BOTH* ORDER AND CHAOS

Deliberative order and creative chaos both have their place—but especially their *time*—however in a seemingly intricate dynamic sequence.

Should Concord let the 'market decide,' instead? They could. Then they could also join the plight of the United States wireless industry, which is several years behind its overseas competitors because of its resolute refusal to agree upon standards serially. The notion of 'picking winners' has a bad name—except in those societies which have come closer to understanding that the economic game of network technology is *both* competitive and cooperative *in dynamic sequence*.

Beside the debacle of the resolutely uncooperative U.S. wireless industry, there is the shining success of the Internet Engineering Task Force (IETF). The IETF has produced one of the most extraordinary runs in the annals of innovation, precisely because it deftly executes the 'change, freeze' cycle countless times to produce the Internet and web. The same behavior is evident also essentially countless times, in the United States, through the working of the many industry-based 'Forums' which spring up when some new technology now requires standardizing. The behavior even has a new name—'open source.'

Competition and the marketplace have their vital place, as the analysis above makes clear. So do concerted action and community. What is key is to appreciate that both must be treated, not statically, but in a dynamic frame that ensconces change.

SOCIALITY OF THE PSYCHE—A CENTRAL PERCEPTION OF THIS APPROACH

Does it make sense to imagine that people will swing back and forth, between competition and consensus? One moment spinning fresh ideas as free spirits, the next shedding that freedom and assuming the cloak of hierarchy, then yet again into the next cycle, and so on? Did it ever make sense to imagine otherwise and expect only competition?

Consider your daily life, in parallel with the following general example: Each person, for a given dimension of life, lives in a world of tighter and wider 'circles of affiliation.' At work, for example, there is membership in the immediate work group, then perhaps a corporate division, then the company itself, even the industry—wider and tighter circles. Moment to moment, a given person will swing from participation in one of these circles, to participation in another. The person may first compete with the work group across the hall, say over budget; then in a next moment s/he may join with that same work group in some common task, perhaps to prosecute a new market opportunity together.[5]

This seemingly complex dance—where competitive and cooperative behavior alternate as blithely as breathing in then out—is built into the human psyche. Core facts of human nature, such as this, must guide the constructs of economics and of course policy as well. Clearly, 'freeze, change' extends this basic human trait into access policy as if hand into a glove.

This perception about the human psyche is one of the bedrocks upon which the approach here proceeds. The chapter now turns to another of its bedrocks.

**The information product—deconstruct it from organization**

The discussion so far has focused on organization and its cyclic alternation. What of the product from this complex dance—the service design that Concord will pick in the thought experiment for example? Call this the information product of the organizational dynamics.

As the community moves into the second, 'freeze' half of the cycle, it will choose a service design to implement the new access policy. In the process, along with the many choices of features unique to the technology, the community will also choose an information architecture that lies somewhere between two extremes: open-layered and vertically integrated.

The extreme ends of this choice can be represented graphically, each with a pyramid.

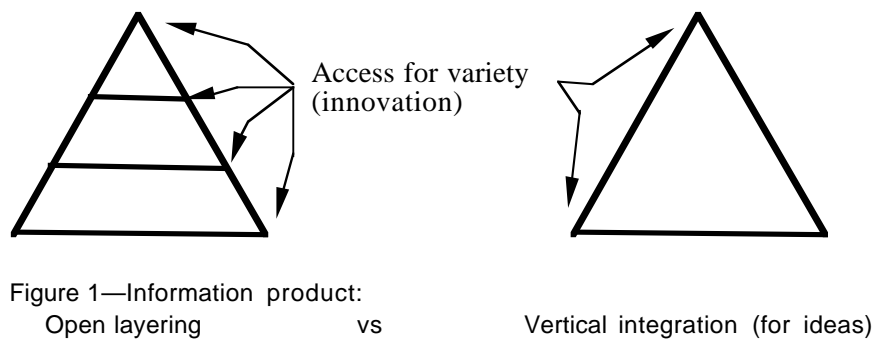

Figure 1—Information product:
   Open layering              vs            Vertical integration (for ideas)

Consider 'open layering' first: These information architectures apply broadly; take Java, from Sun Computing, as an example to describe open layering. Java is an intermediating layer of software. With Java a variety of operating systems, heterogeneous one to the other, can still each produce a common result. So in the 'open' pyramid on the left hand of Figure 1, imagine Java in the middle, intermediate layer. Then the bottom layer of Figure 1 entertains the heterogeneity of, say, a Mac, a PC, and Linux. But the top layer will nonetheless, with Java's intermediation, still produce the same result for users, despite that each is on a different operating system. That is an open layered information architecture.

---

[5] This paragraph is abridged from (Allen 2000, 9).

For the 'vertical integration' choice, take as an example the Apple Macintosh design. From the outset the Mac design has been integrated across its hardware and software.

Vertical integration—for ideas, in the information product—produces higher performance, compared with open layering. For instance Java invariably suffers some performance hit, because of the inevitable overhead necessary to intermediate and translate for the native operating system. So why not always choose the higher performing (information–)vertically integrated design? The reason is compelling, and it informs one-half of the basic architectural choice:  open layering—for the ideas—presents more 'ports' where innovation may potentially enter in the future.

As the community moves through the choices in the 'freeze' phase, it always makes[6] a choice between present performance as against the opportunity for later innovation. Here that is called the choice of architecture for the information product.

Presentday wisdom holds that 'open' is 'good.' Is the community's choice, between open layering and vertical integration (in the ideas), a choice between good and bad? No, decidedly not—it becomes clear that there is a basic tradeoff instead. A *choice* must be made between present performance and future opening to innovations (and *their* potentially higher performance, later).

The earlier discussion of the organizational cycle, between 'change' and freeze,' could be said—with some definitions of 'open' anyway—to have argued that openness in organization is good. For the organizational dynamics, there *is* a 'bad' course to avoid (if not yet discussed explicitly) and that of course diverges from a 'good' path to which the community must hew.

But the choice to be made in the 'freeze' phase is *not* between a 'good' open architecture and 'bad' vertical integration in the ideas—such a description would be nonsensical. Despite current 'wisdom,' there instead must be a (surely complicated and fairly demanding) choice that trades off present performance against future innovation (and perhaps even better performance then).

Only when the dynamics of organization are deconstructed—in our understanding—from the choice of information product, can we see with clarity where openness is mandated (for an organization) versus where it must instead be traded off in a choice (for design). That is the other bedrock, for this approach.

---

[6] If implicitly.

THOUGHT EXPERIMENT, *CONTINUED*—VARIETIES OF 'OPEN'

In the thought experiment, imagine that Concord, as it moves through the 'freeze' phase, chooses fiber-to-the-home, rather than a mix that also brings in fixed wireless local loop. Concord's information architectural choice is vertically integrated then, to enjoy the performance of fiber rather than entertain the future innovation that may arise in both technologies. What about organization?

The conceptual system on which this approach rests has so far addressed only organization for the 'freeze, change' cycle itself. But if extended to organization *for management* after implementation, the impact of deconstructing organization from service design emerges with some prominence. The design choice, in the thought experiment here, is 'integrated.' On the other hand, management for the fiber-to-the-home organization is mandated to be 'open,' if that may prove below to have a surprising twist.

The fiber system itself, just the conduit in other words, will not likely find a competitor.[7] If called a monopoly, the taste is pejorative; if labeled a 'joint effort by the community,' perhaps the taste is more appetizing. Whatever the label, success of the enterprise will have to transpire without the hot breath of competition dogging the town's managers so charged.

Are there models for effective management of what would otherwise be called a monopoly? Some Scandinavian telecoms, as well as others, offer tempting cases. Helsinki Telephone, along with several dozen of its local telco brethren across Finland, have long provided some of the best service, at the lowest prices, and well before there was any suggestion of competition. Telia, née Televerket, in Sweden has done likewise. Singapore's state telecom monopoly was also praised for its efficiency.

THE 'OPEN' ORGANIZATION—A *SERIOUS* CHANGE IN MINDSET

Regardless of whether such a real world model can be agreed, Concord of the thought experiment can reveal the character of *openness for organization* as advocated in this chapter. One cornerstone: responsible use of power aimed at betterment of life in the community. So to choose the next access technology in aid of better communications around the group, the full weight of the community's power is brought to bear. There is no countervailing force against the body politic; clearly to exercise the power of the whole group works only when that power is applied responsibly. Necessary strictures through informal protocols can be intricate. See particularly (Allen 2000, 15–16) for further discussion.[8]

---

[7] The experience with 'overbuilds' in United States cable television systems—an entrant tries to 'build over' an incumbent's system—confirms this empirically. In most overbuilds, one or the other of the two systems drives out its opposite number. The economies of scale with wire-based networks prove to be ineluctable. In a parallel intellectual check, many academics who work in the area, now both in the United States and in Europe, have concluded that 'facilities-based competition' in the case of wire-based local infrastructure is unlikely.

Proscribed by contrast is the opposite—the classic monopolist's use of power to 'drag all the marbles into his/her corner' (the 'bad' course, to be avoided). When a monopolist uses proprietary control of a pivotal standard to club would-be competitors senseless, the behavior-that-must-be-banished has unfortunately been flaunted one more time.

The old, uncomfortable notion of monopoly transforms into real joint effort across the group, for shared benefit. Here is the real change in mindset. Rather than appeal only to greed and self-aggrandizement, the community ethic countenances and preaches sociality *alongside* self-service.

Where are the 'sticks,' the sanctions? How does the group respond to the bully who would spoil the ethic, who would pervert the power of community and subvert it to aggrandize personal ends (and so would become the proscribed monopolist)? The 'spoiler' is 'shunned'—cutoff from the perks of membership in a not-so-subtle nudge to return to the fold.

Rather than greed for never-ending accumulation, service to self is tied with service to the group that *also* sustains each self. Human nature most certainly is not infinitely malleable. But the degree to which these particular views and behaviors can be swayed by early socialization is probably significant. To an important extent, the community gets what it preaches: self or self-plus.

Thus this mindset change is in the hands of the community. However, such a change in mindset may be even more challenging to put across—the adoption of a dynamic outlook may be a romp by comparison …

In the thought experiment Concord proactively chose an information product that is 'vertically integrated.' But its [management] organization is 'open' by mandate of the general analysis here, despite being a nominal monopoly. The terms of the thought experiment were chosen of course to highlight the contrast, to emphasize how important it is to deconstruct—to separate—ideas about organization from those for service design. Only then is it possible to be clear about the correct, and markedly different, paths for each.

This is another bedrock, if we are to get right the whole of policy for access. There are really different notions of openness, to be applied separately. On the one side for the dynamics of community organization and the evolution of policy, there is a good and bad, a right and wrong where open is the desideratum. On the other side for the informational product of the community's work and the design of a service, open is just

---

[8] For that matter, see most introductory civics texts for a parallel discussion …

at one end of a spectrum where there is a choice to be made, a point to be selected between open on one end and integrated at the other.

**Who is us?**

In the United States, a small town such as Concord is doing well if it achieves a sense of cohesive community. Such are the offshoots of multi-cultural society. Generally cohesion extends just to the individual, small 'pots' where cultural commonalties may be found. Contrasts abound around the world, however, particularly where social democracy is more the norm. Japan is just one case.

A small town in Japan may proudly offer a cuisine distinct from all its neighboring towns; the local dialect may even be at some variance from NHK Japanese. At the same time Japan as a society has 'rolled up' its individual, distinct pots and connected them into a wider, larger sense of community. As a result, there is *also* an 'us' across essentially the whole population of the archipelago.

The impact on access policy—at least what the impact might be—is worth a look. A larger community makes possible what some will think of as a tops-down policy. As of late 1999, NTT had installed fiber-to-the-curb for about twenty percent of the main lines in Osaka and Tokyo, the two largest urban centers.[9] That is millions and millions of main lines, already upgraded to undeniably broadband access, effectively in one fell swoop. And the march continues, ultimately for the rest of the country.

There are similar cases, from other cultures; Japan is only one example among a number.

Profoundly, the size of 'us' directly dictates the possibilities for access. If the United States could 'roll up' its 'pots' to be orders of magnitude greater than say the 17,000 population of a Concord, the prospects for access policy then change in direct proportion.[10] The United States might then transcend its bottoms-up approach. Whether on the other hand 'it wants to do so' does go to the heart of change in the mindset.

**Will it work?**

Despite the world examples counter to the U.S. situation, community does not scale well. Particularly that is the case in the face of multi-culturality. (Though whether the number of multi-cultural societies will grow and predominate is still an open question.[11]) But as highways, fast jetliners, and even faster fiber optics have put people from faraway places in almost instant touch with each other, the geographic basis for traditional nuclear community is sorely tried.

---

[9] Personal conversation with Mr. Kiyoshi Isozaki, President, Chief Executive Officer, NTT Communications, Telecom 99, October, Geneva.

[10] How much each individual's sense of 'the group' overlaps with that of others is a separate question. For the analysis here, it is sufficient that while overlap is never exact the effect is sufficient for the shared sense of community. Indeed, that can be observed universally, for some size group.

[11] Despite the rhetoric for globalism and a universal melting pot.

The access policy envisioned here depends upon precisely the sense of shared ties and particularly the informal style of 'regulation' that is second nature to any close-knit and well-oiled community. Could a United States, despite all and despite 'modern' cosmopolitan urban life, learn how to roll up its pots and make the connections for a larger community?

Regardless of the answer, a new *joint* standard is now clear. To date "market failure" has been the test. Now that is joined—in equal measure—by "community failure."

But the spirit of this analysis turns from the negatives of failure—it puts prospects and policy in positive terms. Policy is founded on a realistic bedrock, the sociality of the psyche. Another bedrock treats open as a normative guide for organization; but it understands that the organization dynamic chooses a point between open (different from the 'social' open) and integrated, for design.

This positive access policy can alternate between concerted action and competition together—deliberate order and creative chaos *together* are the new dynamic mantra.

NB: Open links to access URL's, if viewing electronically.